\documentclass{raa}

\usepackage{graphicx,times}             
\input{epsf.sty}

\newcommand{\mdot}{\mbox{$\dot{M}$}}

\newcommand{\lsim}{\raisebox{-.4ex}{$\stackrel{<}{\scriptstyle \sim}$}}

\begin{document}

\title{Transition from radiatively inefficient to cooling dominated
phase in two temperature accretion discs around black holes
}

\author{Monika Sinha \and  S. R. Rajesh \and Banibrata
Mukhopadhyay} 

\institute{Astronomy and Astrophysics Program, Department of Physics,
Indian Institute of Science, Bangalore 560012, India; 
{\it msinha@physics.iisc.ernet.in, rajesh@physics.iisc.ernet.in, 
bm@physics.iisc.ernet.in} \\ 
}


\abstract{
We investigate the transition of a radiatively inefficient phase of 
a viscous two temperature accreting flow to a cooling dominated phase
and vice versa around black holes. Based on a global sub-Keplerian
accretion disc model in steady state, including explicit cooling processes
self-consistently, we show that general advective accretion
flow passes through various phases
during its infall towards a black hole.
Bremsstrahlung, synchrotron and inverse Comptonization of soft photons
are considered as possible cooling mechanisms.
Hence the flow governs a  much lower electron temperature $\sim 10^8-10^{9.5}$K
compared to the hot protons of temperature $\sim 10^{10.2}-10^{11.8}$K
in the range of the accretion rate in Eddington units
$0.01\lsim\mdot\lsim 100$. Therefore, the solutions may potentially
explain the hard X-rays and the $\gamma$-rays emitted from AGNs and 
X-ray binaries.  We finally compare the solutions for two different 
regimes of viscosity and conclude that a weakly viscous flow is 
expected to be cooling dominated compared to its highly viscous
counterpart which is radiatively inefficient. The flow is successfully
able to reproduce the observed luminosities of the under-fed AGNs and
quasars (e.g. Sgr~$A^*$), ultra-luminous X-ray sources (e.g. SS433), as
well as the highly luminous AGNs and ultra-luminous quasars (e.g. PKS~0743-67)
at different combinations of the mass accretion rate and ratio of
specific heats.
}

\authorrunning{Sinha, Rajesh \& Mukhopadhyay}

\titlerunning{Transition between radiatively inefficient and cooling 
dominated phases in accretion discs}

\maketitle


\section{Introduction}

The observed hard X-rays from, e.g., Cyg~X-1 can not be explained
(Lightman \& Shapiro 1975) by the model of cool
Keplerian accretion disc (Pringle \& Rees 1972;
Shakura \& Sunyaev 1973; Novikov \& Thorne 1973).
Indeed Eardley \& Lightman (1975) found
that a Keplerian disc is unstable due to thermal and viscous effects
with constant viscosity parameter $\alpha$ (Shakura \& Sunyaev 1973)
which was later shown by Eggum, Coroniti \& Katz (1985) by numerical simulations.
Since then, the idea of two component accretion disc
started floating around. For example,
Muchotrzeb \& Paczy\'nski (1982) introduced the idea of sub-Keplerian,
transonic accretion,
which was later improved by other authors (Chakrabarti 1989, 1996;
Mukhopadhyay 2003).

Shapiro, Lightman \& Eardley (1976) introduced a two temperature Keplerian
accretion disc at a low mass accretion rate which is
significantly hotter than the single temperature Keplerian disc of Shakura
\& Sunyaev (1973).
Various states of Cyg~X-1 could be well explained by this model (e.g. Melia \& Misra 1993).
However, in this model, solutions
appear thermally unstable.
Narayan \& Popham (1993) and subsequently Narayan \& Yi (1995)
introduced advection to stabilize the system. However, this model,
with inefficient cooling mechanisms, could explain only a particular
class of hot systems.
Moreover, the model kept the electron heating decoupled
from the disc hydrodynamics.
On the other hand, Chakrabarti \& Titarchuk (1995) and
later Mandal \& Chakrabarti (2005) modeled similar kind of flows
emphasizing possible formation of shock and its consequences
therein. However, they also did not consider the effect of electron heating
self-consistently into the hydrodynamic equation, and thus
the hydrodynamic
quantities remain decoupled from the rate of electron heating.

In the present paper, we model a self-consistent accretion disc in
two temperature transonic sub-Keplerian regime. Considering all
the hydrodynamic equations of the disc along with thermal components
we solve the coupled set of equations self-consistently.
We investigate
switching over the flow during infall, from the radiatively inefficient 
nature, e.g. ADAF (Narayan \& Yi 1994), to
general advective paradigm and then to cooling dominated phase and vice versa.
In order to apply the model to explain observations,
we focus on the ultra-luminous X-ray (ULX) sources (e.g. SS433),
under-luminous AGNs and quasars (e.g. Sgr~$A^{*}$) and
ultra-luminous quasars and highly luminous AGNs
(e.g. PKS~0743-67), while the first set of objects
is likely to be the ``radiation trapped'' accretion disc.

In the next section, we discuss the model equations governing the
system and the procedure to solve them. In \S3 and \S4, we describe
the two temperature accretion disc around stellar mass
and supermassive black holes respectively.
Section 5 compares the disc flow of low Shakura-Sunyaev (1973) $\alpha$ with
that of high $\alpha$. Then we summarize the results in \S6 with implications.

\section{Formalism}

We set five coupled differential equations to
describe the laws of conservation in the sub-Keplerian advective accretion disc,
which is presumably optically thin.
All the variables are expressed throughout in dimensionless units, unless stated otherwise.
The radial velocity $v$ and sound speed $c_s$ are expressed
in units of light speed $c$, the specific angular momentum $\lambda$
in $GM/c$, where $G$ is the Newton's gravitational constant and $M$ is the
mass of the black hole,
expressed in units of solar mass $M_\odot$, the
radial coordinate $x$ in units of $GM/c^2$, the density $\rho$ and the total pressure $P$
accordingly. The disc fluid, which behaves as an (almost) noninteracting gas, 
consists of ions and electrons, apart from radiation.

\subsection{Conservation laws }
(a) Mass transfer:
\begin{eqnarray}
\frac{1}{x} \frac{\partial}{\partial x} (x \rho v)  =  0,\,\,
{\rm and\,\,thus}\,\,
\mdot\,=\,-4\pi x \Sigma v,
\label{mass}
\end{eqnarray}
where the surface density
\begin{eqnarray}
\Sigma \,= \,I_n\, \rho_{eq}\, h(x),\,
I_n \,= \,(2^{n} n!)^{2}/(2n+1)!\,\, {\rm (Matsumoto\hskip0.2cm et\hskip0.2cm al.\hskip0.2cm 1984)},
\label{matsu}
\end{eqnarray}
$\rho_{eq}$ is density at the equatorial plane,
half-thickness of the disc
\begin{eqnarray}
h(x) \,= \,c_s x^{1/2} F^{-1/2},
\label{thik}
\end{eqnarray}
$F$ is the magnitude of gravitational force.

(b) Radial momentum balance:
\begin{eqnarray}
v \frac{dv}{dx} \,+ \,\frac{1}{\rho} \frac{dP}{dx} \,- \,\frac{\lambda^{2}}{x^3} \,+ \,F \, = \, 0
\label{rad}
\end{eqnarray}
when following Mukhopadhyay (2002) the gravitational force corresponding to the 
pseudo-Newtonian potential
\begin{eqnarray}
F=\frac{(x^2 - 2a\sqrt{x} +a^2)^2}{x^3[\sqrt{x}(x - 2) +a]^2},
\end{eqnarray}
where $a$ is the Kerr parameter, which, for simplicity, 
is chosen to be zero (Schwarzschild black hole)
for the present purpose.
Following Ghosh \& Mukhopadhyay (2009), we also define 
\begin{eqnarray}
\beta=\frac{{\rm gas\hskip0.2cm pressure\hskip0.2cm} P_{gas}}
{{\rm total\hskip0.2cm pressure\hskip0.2cm} P}\,=\frac{6\gamma-8}{3(\gamma-1)}
\end{eqnarray}
where $\gamma$ is the ratio of specific heats given by
$4/3\le\gamma\le5/3$, $P_{gas}=P_i\hskip0.2cm({\rm ion\hskip0.2cm pressure})
+P_e\hskip0.2cm({\rm electron\hskip0.2cm pressure})$, such that
\begin{eqnarray}
P \,= \,\frac{\rho}{\beta\,c^2} \left(\frac{kT_{i}}{\mu_{i} m_{i}} \,+
\,\frac{kT_{e}}{\mu_{e} m_{i}}\right) \,= \,\rho c^{2}_{s},
\label{ptot}
\end{eqnarray}
where $T_{i}$, $T_{e}$ are respectively the ion and electron temperatures in Kelvin,
$m_i$ is the mass of a proton in gm, $\mu_i$ and $\mu_e$
respectively are the corresponding effective molecular weight, $k$ the Boltzmann constant.
Note that for $\gamma=4/3$, $\beta=0$; pure radiation flow, and for $\gamma=5/3$, 
$\beta=1$;  pure gas flow. 

(c) Azimuthal momentum balance:

\begin{eqnarray}
v \frac{d \lambda}{dx} \,= \,\frac{1}{\Sigma x} \frac{d}{dx}\left(x^2 |W_{x \phi}|
\right),
\label{az}
\end{eqnarray}
where following Mukhopadhyay \& Ghosh (2003) the shearing stress is given by
\begin{eqnarray}
W_{x \phi}  \,= \,- \alpha  \left(I_{n+1} P_{eq} \,+ \,I_n v^2 \rho_{eq} \right)h(x),
\end{eqnarray}
where $\alpha$ is the dimensionless viscosity parameter and $P_{eq}$ and $\rho_{eq}$ are
the pressure and density respectively
at the equatorial plane which will be assumed to be the same as general
$P$ and $\rho$ respectively
in obtaining solutions.

(d) Energy production rate:

\begin{eqnarray}
\frac{v h(x)}{\Gamma_{3} - 1} \left(\frac{dP}{dx} \,- \,\Gamma_{1} \frac{P}{\rho} \frac{d \rho}{dx}\right) \,= \,Q^{+} \,- \,Q_{ie},
\label{eni}
\end{eqnarray}
where following Mukhopadhyay \& Ghosh (2003) the heat generated by viscous dissipation
\begin{eqnarray}
Q^+ \,= \, \alpha (I_{n+1} P \,+ \,I_n v^2 \rho )h(x) \frac {d \lambda}{dx},
\label{qvis}
\end{eqnarray}
when the Coulomb coupling, written in dimensionless form  
(Bisnovatyi-Kogan \& Lovelace 2000) $Q_{ie}$ is given in the 
dimensionful unit as
\begin{eqnarray}
q_{ie} \, = \,\frac{8 (2 \pi)^{1/2}e^4 n_i n_e}{m_i m_e}\left(\frac{T_e}{m_e} \,+ \, \frac{T_i}{m_i}\right)^{-3/2} \ln (\Lambda) \ \left(T_i \, - \,T_e \right)\,\,{\rm erg/cm^3/sec},
\label{qie}
\end{eqnarray}
where $n_i$ and $n_e$ respectively denote number densities of ion and electron,
$e$ the electron charge,
ln($\Lambda$) the Coulomb logarithm.
We also define 
\begin{eqnarray}
\Gamma_{3} &= &1 \,+ \,\frac{\Gamma_{1} \,- \,\beta}{4 \,- \,3 \beta},\\
\Gamma_{1} &= &\,\beta \,+ \,\frac{(4 \,- \,3 \beta)^{2}(\gamma \,- \,1)}{\beta+\,12(\gamma \,- \,1)(1 \,- \,\beta)}.
\label{gam}
\end{eqnarray}

(e) Energy radiation rate:

\begin{eqnarray}
\frac{v h(x)}{\Gamma_{3} - 1} \left(\frac{dP_{e}}{dx} \,- \,\Gamma_{1} \frac{P_{e}}{\rho} \frac{d \rho}{dx}\right) \,= \,Q_{ie} \,-\,Q^{-},
\label{ene}
\end{eqnarray}
where the total heat radiated away ($Q^-$) by the
bremsstrahlung ($q_{br}$), synchrotron ($q_{syn}$) processes and inverse Comptonization
($q_{comp}$) due to soft synchrotron photons is given in dimensionful form as
(Narayan \& Yi 1995; Mandal \& Chakrabarti 2005)
\begin{eqnarray}
q^-=q_{br}+q_{syn}+q_{comp},
\label{qm}
\end{eqnarray}
where
\begin{eqnarray}
\nonumber
q_{br} &= &1.4 \times 10^{-27} \ n_e\,n_i T_e^{1/2}\,(1+4.4\times 10^{-10} T_e)\,
\,\,{\rm erg/cm^3/sec},\\
\nonumber
q_{syn} & = &\frac{2 \pi}{3 c^2} kT_e \, \frac{\nu_a^{3}}{R}\,\,\,{\rm erg/cm^3/sec},
\,\,\,R=x\,GM/c^2,\\
\nonumber
q_{comp} &=& {\cal F}\,q_{syn},\,\,\ {\cal F} = \eta _{1}\,
\left(1 \,- \,\left(\frac{x_{a}}{3 \theta _{e}}\right)^{\eta _{2}} \right),\,\,\,
\eta _{1} = \frac{p(A-1)}{(1-pA)},\,\,\
p \,= \,1 \,- \,\exp(- \tau _{es}),\\
A &= &1 \,+ \,4 \theta _{e} + \,16 \theta^{2}_{e},\,\,\,\theta _{e} \,=
\,kT_{e}/m_{e} c^{2},\,\,\
\eta _{2} = -\, 1 \,- \,\frac{ln(p)}{ln(A)},\,\,\,x_{a} \,= \,h \nu_{a}/m_{e} c^{2},
\label{qvari}
\end{eqnarray}
when $\tau_{es}$ is the scattering optical depth given by
\begin{eqnarray}
\tau_{es}=\kappa_{es}\rho\,h
\end{eqnarray}
where $\kappa_{es}=0.38$ cm$^2$/gm 
and $\nu_a$ is the
synchrotron self-absorption cut off frequency. However, the total optical
depth should include the effects of absorption due to nonthermal processes.
Therefore, effective optical depth is given by
\begin{eqnarray}
\tau_{eff}\simeq\sqrt{\tau_{es}\,\tau_{abs}}
\end{eqnarray}
where approximately 
$\tau_{abs}\simeq 6\times 10^{23}\,\rho^2\,T_e^{-7/2}\,h$.

Now, combining all the equations we obtain
\begin{eqnarray}
\frac{dv}{dx} \,= \,\frac{N(x,v,\lambda,c_{s},T_{e})}{D(v,c_{s})},
\label{dvdx}
\end{eqnarray}
where
\begin{eqnarray}
\nonumber
N \,= \,\frac{\Gamma_1 \,+ \,1}{\Gamma_3 \,- \,1} v^{2} c_s J\,- \,
\frac{ \alpha^{2}  c_s}{x} H\left( \frac{I_{n+1}}{I_n} c_s^{2} \,+ \,v^{2} \right) \,- \,
\alpha^{2} \frac{I_{n+1}}{I_n} 2H J\,+ \,
  \frac{\Gamma_1 \,- \,1}{\Gamma_3 \,- \,1} v^{2} c_s^{3} G \,
+ \,\alpha H \left(\frac{2\lambda v c_s}{x^{2}} \right) \\
\nonumber
+ \,\frac{4\pi Q_{ie}}{\mdot}v^{2} c_s^{2} x^{3/2} F^{-1/2},\\
\label{num}
\end{eqnarray}
\begin{eqnarray}
D \,= \,\frac{1 \,- \,\Gamma_1 }{\Gamma_3 \,- \,1} c_s^{3} v \,+ \,2 \alpha c_s \frac{I_{n+1}}{I_n}H \left(\frac{c_s^{2}}{v} \,- \,v \right) \,+ \,\frac{\Gamma_1 \,+ \,1}{\Gamma_3 \,- \,1} v^{2} c_s^{2} \left(v \,- \,\frac{c_s^{2}}{v}\right) \,+ \,\alpha^{2} v H \left(\frac{H}{v}\right)
\label{den}
\end{eqnarray}
and
\begin{eqnarray}
G \,= \,\left(\frac{3}{2x} \,- \,\frac{1}{2F}\frac{dF}{dx} \right),\,\,
H \,= \,\left( I_{n+1} c_s^{2} \,+ \,I_n v^{2} \right),\,\,
J \,= \,\left( c_s^{2} G \,+ \,\frac{\lambda^{2}}{x^{3}} \,- \,F\right).
\end{eqnarray}

Finally combining eqns. (\ref{rad}), (\ref{az}) and (\ref{ene}) we obtain
\begin{eqnarray}
\frac{d c_{s}}{dx} \,= \,\left(\frac{c_{s}}{v} - \frac{v}{c_{s}}\right) \frac{dv}{dx} \,+
\,\frac{J}{c_{s}},
\label{dcsdx}
\end{eqnarray}
\begin{eqnarray}
\frac{d \lambda}{dx} \,= \,\left(\frac{2 \alpha x}{v c_{s}} \frac{I_{n+1}}{I_{n}}
\left(\frac{c_{s}^{3}}{v}-v c_{s}\right)+\alpha x\right) \frac{dv}{dx} \,+
\,\left(\frac{c_{s}^{2}-2x \alpha J}{c_{s}}+v\right),
\label{dldx}
\end{eqnarray}
\begin{eqnarray}
\frac{dT_{e}}{dx} \,= \,(1- \Gamma _{1})T_{e} \frac{v}{c_{s}^{2}} \frac{dv}{dx} \,+ \,(1-
\Gamma _{1})T_{e}\left(\frac{J}{c_{s}^{2}}+G\right) \,+ \,\frac{(\Gamma_{3}-1)4
\pi}{\mdot}\frac{c_{s} x^{3/2}}{F^{1/2}} \left(Q^{ie}-Q^{-}\right).
\label{dtedx}
\end{eqnarray}
Now following the procedure adopted in the  previous works (e.g. Chakrabarti 1996, Mukhopadhyay 2003, Mukhopadhyay \&
Ghosh 2003) we solve eqns. (\ref{dvdx}), (\ref{dcsdx}), (\ref{dldx}), (\ref{dtedx}) 
for $v$, $c_s$, $\lambda$, $T_e$.

There is a possibility of convective instability in advective flows as the entropy increases inwards
(e.g. Narayan \& Yi 1994,
Chakrabarti 1996). This may help in explaining transport as well, as proposed by Narayan \& Yi (1994).
When the square of effective frequency 
\begin{eqnarray}
\nu_{eff}^{2} \,= \,\nu_{\rm BV}^{2} \,+ \,\nu_r^{2}\,<\,0
\end{eqnarray}
dynamical convective instability arises, where
$\nu_{\rm BV}$ is the Brunt-V\"ais\"al\"a frequency and $\nu_r$ the radial epicyclic frequency
given by
\begin{eqnarray}
\nu_{\rm BV}^{2} \,= \,-\frac{1}{\rho} \frac{dP}{dx} \frac{d}{dx}ln\left(\frac{P^{1/
\gamma}}{\rho}\right),\,\,\,\,\ \nu_r^{2}\,=\,\frac{2\lambda}{x^3}\,\frac{d\lambda}{dx}.
\end{eqnarray}

\subsection{Solution procedure}
As previous works (e.g. Chakrabarti 1996, Mukhopadhyay 2003, Mukhopadhyay \& Ghosh 2003), 
in order to obtain the steady state solution we primarily need
to find out the self-consistent value of the sonic/critical radius $x_c$ and the corresponding
specific angular momentum $\lambda_c$ of the flow. For the present purpose of
a two temperature flow, at $x_c$ the electron temperature $T_{ec}$ also needs to be
specified. Note that the set of values $x_c, \lambda_c, T_{ec}$ has to be adjusted appropriately
to obtain self-consistent solution connecting the
outer boundary to the black hole event horizon through $x_c$. 
Depending on the type of accreting
system to model, we then have to specify the related
inputs: $\mdot$, $M$ and $\gamma$. 
Importantly, unlike former works (e.g. Chakrabarti \& Titarchuk 1995,
Chakrabarti 1996, Mukhopadhyay \& Ghosh 2003) here $x_c$ changes with 
the change of $\mdot$, which is very natural
because the various cooling processes considered here explicitly depend on $\mdot$.
Finally, we have
to solve the Eqn. (\ref{dvdx}) from $x_c$ to the
black hole event horizon, and then outwards upto the transition
radius $x_o$ where the disc
deviates from the Keplerian to the sub-Keplerian regime such that $\lambda/\lambda_K=1$
($\lambda_K$ being the specific angular momentum of the
Keplerian part of the disc).

\section{Disc flows around stellar mass black holes}

We concentrate on the super-Eddington accretor which presumably is the
case of ultra-luminous X-ray binaries.
We mainly intend to understand the explicit dependence of the disc
properties on the cooling processes and then the variation of the cooling
efficiency $f$ with the disc radii.
Note that $f$ is defined to be the ratio of the energy advected by the
flow to the energy dissipated; $f\rightarrow 1$ for
the advection dominated accretion flow (in short ADAF;
Narayan \& Yi 1994, 1995) and $f<1$
for the general advective accretion flow (in short GAAF; Chakrabarti 1996,
Mukhopadhyay 2003, Mukhopadhyay \& Ghosh 2003).
Far away from the black hole the disc becomes (or tends to become) of one
temperature when the gravitational power
is weaker and hence the angular momentum profile remains Keplerian 
in the presence of efficient cooling.

Note that the ``radiation trapped'' accretion disc can be attributed to
the radiatively driven outflow and jet which 
is likely to occur when the accretion rate is super-Eddington
(Lovelace, Romanova \& Newman 1994, Begelman, King \& Pringle 2006, 
Fabbiano 2004, Ghosh \& Mukhopadhyay 2009),
as seen in the ULX sources such as SS433 (with luminosity
$\sim 10^{40}$ erg/s or so; Fabrika 2004). In order to describe such sources,
we consider $\mdot=10$. 
Throughout we express $\mdot$ in units of Eddington limit. The set of input parameters
used for this case is given in Table 1. 
However, a detailed work of the two temperature
viscous accretion flows in the possible range of accretion rates: $0.01\lsim\mdot\lsim100$, 
around rotating stellar mass black holes is under
preparation (Rajesh \& Mukhopadhyay 2009)

\begin{figure}
\centering
\includegraphics[angle=-90,width=0.90\columnwidth]{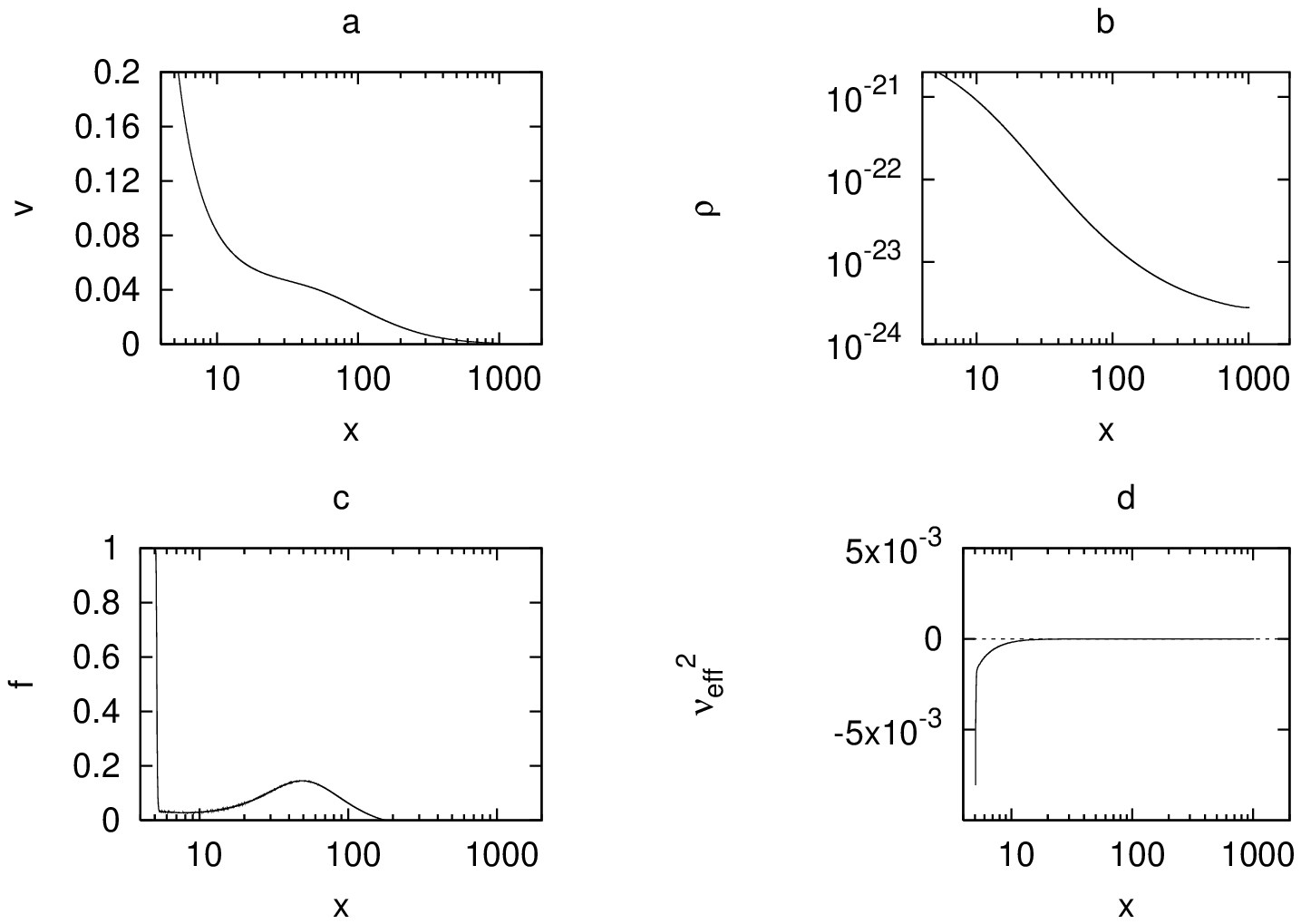}
\caption{
Variation of dimensionless (a) radial velocity, (b) density,
(c) cooling factor,
(d) square of convective frequency, as functions of radial
coordinate for $\mdot=10$.
Other parameters are $\alpha = 0.01$, $M=10$, $\gamma=1.345$; see Table 1 for details.
 }
\label{figsth0}
\end{figure}

\begin{figure}
\centering
\includegraphics[angle=-90,width=0.90\columnwidth]{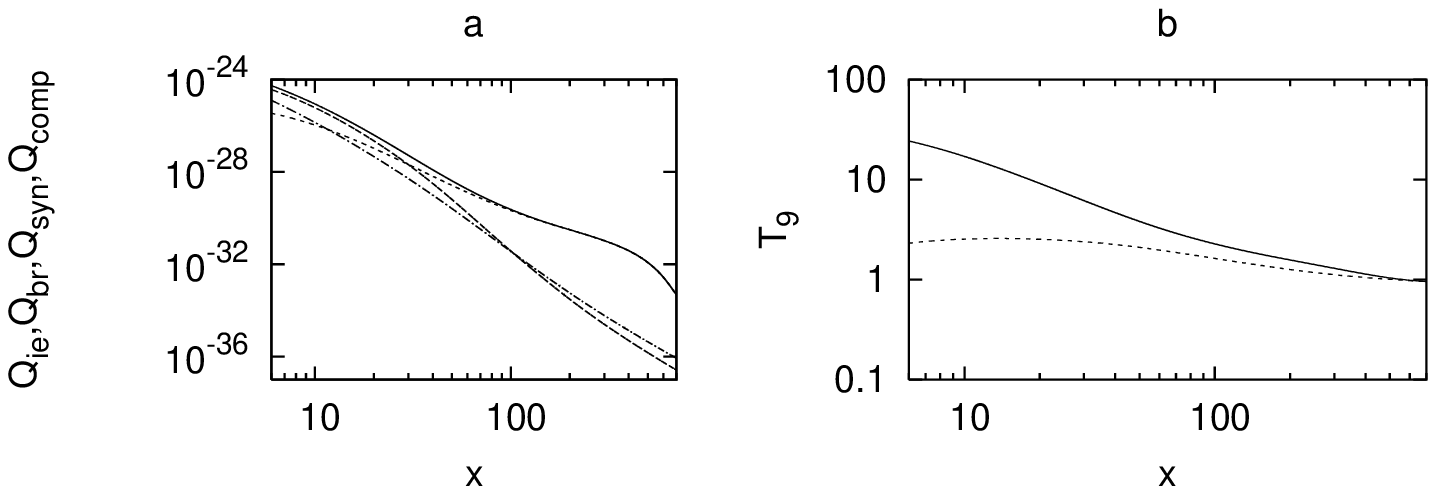}
\caption{
Variation of (a) dimensionless energy of Coulomb coupling (solid line),
bremsstrahlung (dotted line), synchrotron (dash-dotted line), inverse Comptonization due
to synchrotron photon (dashed line) processes in logarithmic scale,
(b) corresponding
ion (solid) and electron (dotted) temperatures in units of $10^9$K,
as functions of radial coordinate for $\mdot=10$.
Other parameters are $\alpha = 0.01$, $M=10$, $\gamma=1.345$; see Table 1 for details.
}
\label{figsth0t}
\end{figure}

A high mass accretion rate renders density to be very high which is very favourable
for various cooling mechanisms. Naturally this results in
$f$ being low which finally affects the two temperature nature. The profile of
velocity shown in
Fig. \ref{figsth0}a clearly indicates a centrifugal barrier at around $x\sim 30$. 
However, further out, at $x\sim 50$, $f$ increases (see Fig. \ref{figsth0}c)
as the energy radiated due to
the bremsstrahlung process is weaker than the energy transferred from
protons to electrons through the Coulomb coupling (see Fig. \ref{figsth0t}).
Subsequently, the
synchrotron process becomes dominant (see Fig. \ref{figsth0t}),
causing $f\rightarrow 0$. However, the presence of strong advection
near the black hole does not allow the flow to
radiate efficiently rendering $f\rightarrow 1$ again. This also
leads to marginal convective instability only at $x<10$, as is evident
from Fig. \ref{figsth0}d (see, however, Narayan, Igumenshchev \& Abramowicz 2000,
Quataert \& Gruzinov 2000). 

Figure \ref{figsth0t} shows the variation of cooling processes and the
corresponding temperature profiles with the radial coordinate. 
At the transition radius the disc remains of one
temperature (see Fig. \ref{figsth0t}). However, a strong two temperature nature
appears when the flow advances with a sub-Keplerian angular
momentum. This is because far away
the electrons and ions are in thermal
equilibrium, around the transition radius,
particularly at a high $\mdot$. As matter infalls through the sub-Keplerian regime,
the ions become hotter
rendering the ion-electron Coulomb collisions weaker. However, the 
electrons cool down via processes like bremsstrahlung, synchrotron 
emissions etc. keeping their temperature roughly constant upto 
very inner disc. As a result, while far away from the black hole the 
flow is in the cooling dominated phase, it transits (or tends to 
transit) to a radiatively inefficient phase, e.g. ADAF, close to the 
black hole.

\section{Disc flows around supermassive black holes}

Here we concentrate on two extreme regimes:
(1) sub-Eddington limit of accretion with $\mdot=0.01$, which is
presumably the case of under-luminous AGNs, (2) super-Eddington accretion
with $\mdot=10$,
which presumably mimics ultra-luminous quasars and highly luminous AGNs.
However, a detailed work of two temperature
viscous accretion flows in the possible range of accretion rates: 
$0.00001\lsim\mdot\lsim100$, around rotating supermassive black holes 
is under preparation (Rajesh \& Mukhopadhyay 2009) 

\subsection{Sub-Eddington accretors}

The under-luminous AGNs and quasars (e.g. Sgr~$A^{*}$)
could be described by the advection dominated model,
where the flow is likely to be substantially sub-critical/sub-Eddington
with a very low luminosity ($\lsim 10^{35}$ erg/s).
Therefore the present case of $\mdot\lsim 0.01$
could be an appropriate model for describing under-luminous
sources.
The parameters for the model case described here are given in Table 1.

\begin{figure}
\centering
\includegraphics[angle=-90,width=0.90\columnwidth]{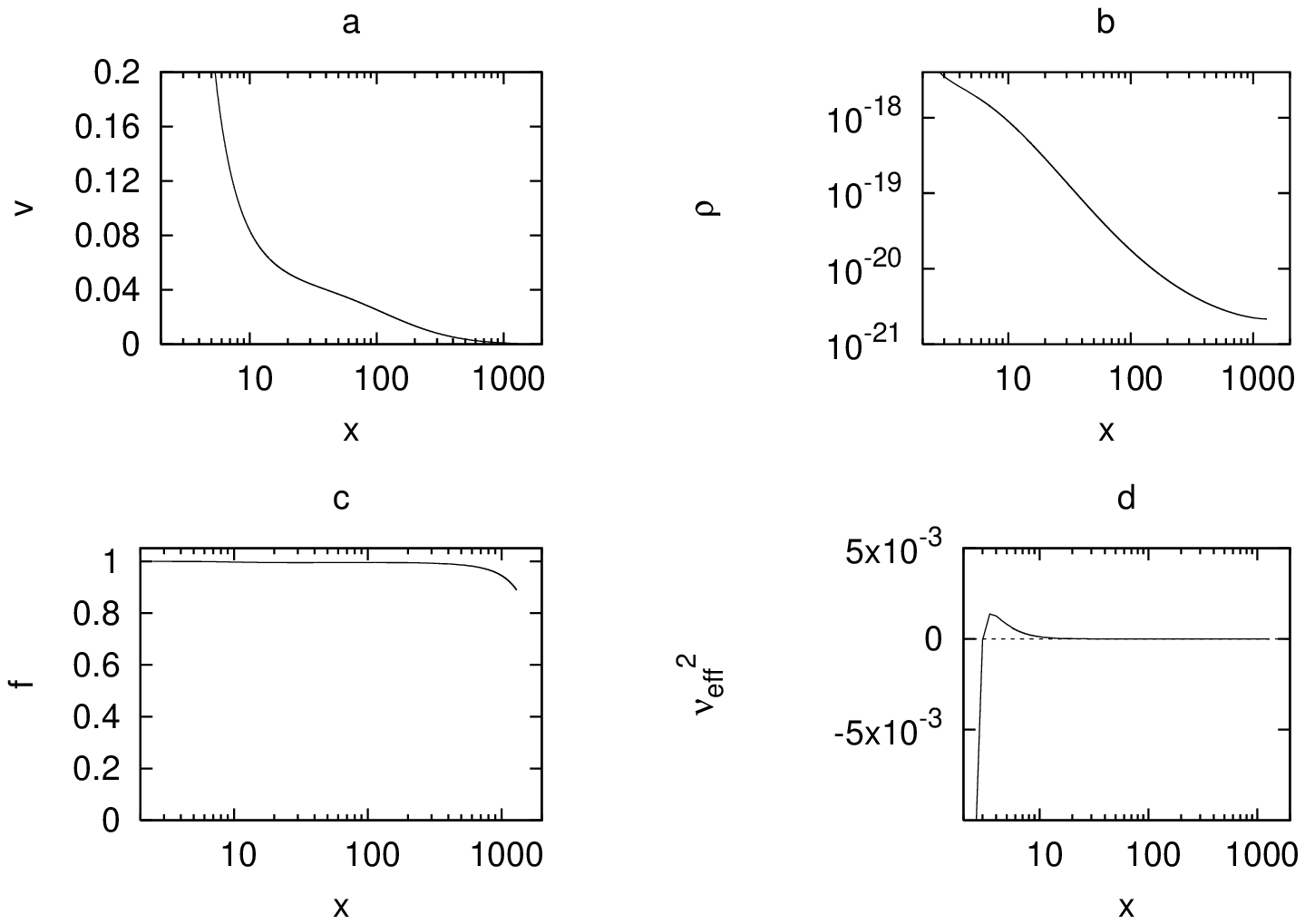}
\caption{
Variation of dimensionless (a) radial velocity, (b) density,
(c) cooling factor,
(d) square of convective frequency, as functions of radial
coordinate for $\mdot=0.01$.
Other parameters are $\alpha = 0.01$, $M=10^7$, $\gamma=1.5$; see Table 1 for details.
 }
\label{figsul0}
\end{figure}

Naturally the density of the disc around a supermassive black hole is
much lower compared to that around a stellar mass black hole.
Therefore, the cooling
processes, particularly the bremsstrahlung radiation which is only density
dependent,  become inefficient leading to a high $f$.
However, Fig. \ref{figsul0}a shows that the velocity profile is 
similar to that around a stellar mass black hole.
From Fig. \ref{figsul0}c we see that $f\rightarrow 1$ in most of
the sub-Keplerian regime.
Close to the black hole
there is a possible convective instability as shown in Fig. \ref{figsul0}d.
This is due to a strong advection of matter. 

\begin{figure}
\centering
\includegraphics[angle=-90,width=0.90\columnwidth]{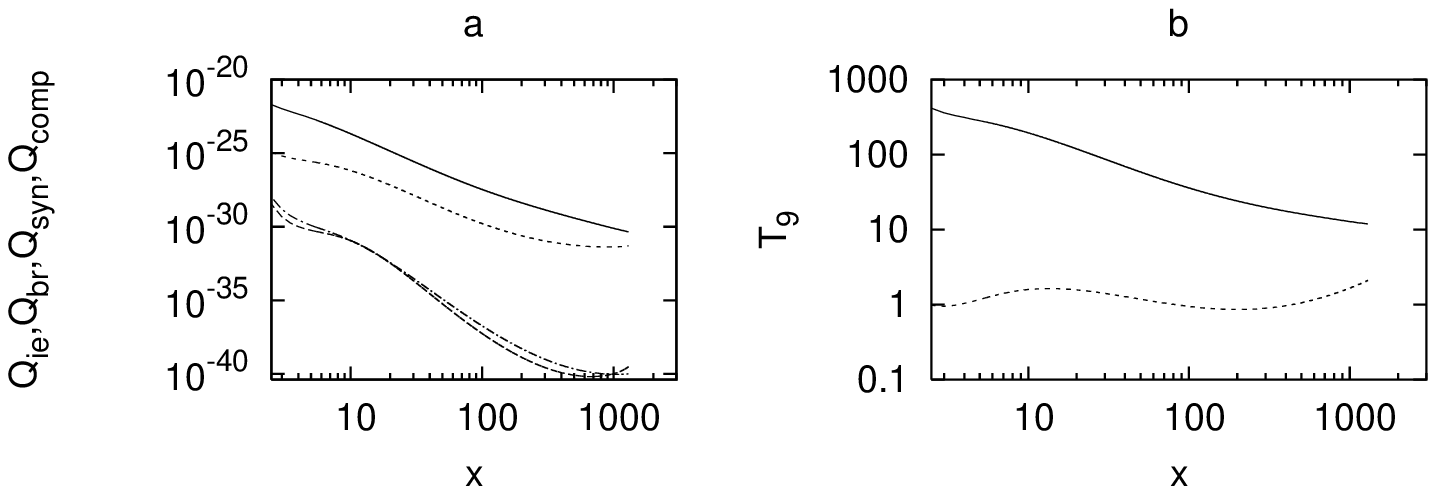}
\caption{
Variation of (a) dimensionless energy of Coulomb coupling (solid line),
bremsstrahlung (dotted line), synchrotron (dash-dotted line), inverse Comptonization due
to synchrotron photon (dashed line) processes in logarithmic scale,
(b) corresponding
ion (solid) and electron (dotted) temperatures in units of $10^9$K,
as functions of radial coordinate for $\mdot=0.01$.
Other parameters are $\alpha = 0.01$, $M=10^7$, $\gamma=1.5$; see Table 1 for details.
}
\label{figsul0t}
\end{figure}

Figure \ref{figsul0t} shows the variation of cooling processes and
the corresponding temperature profiles with the radial coordinate. A low
$\mdot$ corresponds to a radiatively inefficient hot 
two temperature Keplerian-sub-Keplerian transition region.
However, a point to be noted is that unlike stellar mass black holes, only the
bremsstrahlung radiation is effective in cooling the flow around a
supermassive black hole. This is due to a low magnetic field in the
disc around a supermassive black hole rendering synchrotron radiation insignificant.

\subsection{Super-Eddington accretors}

The highly luminous AGNs and ultra-luminous quasars with radio jet
(e.g. PKS~0743-67; Punsly \& Tingay 2005), possibly in ULIRGs
(Genzel et al. 1998) and narrow-line Seyfert 1 galaxies
(e.g. Mineshige et al. 2000), are likely to be ultra-luminous accretors
with a high kinetic luminosity ($\sim 10^{46} - 10^{49}$ erg/s).
Therefore, the parameter set given in the last row of Table 1 
could be appropriate to describe such sources.

\begin{figure}
\centering
\includegraphics[angle=-90,width=0.90\columnwidth]{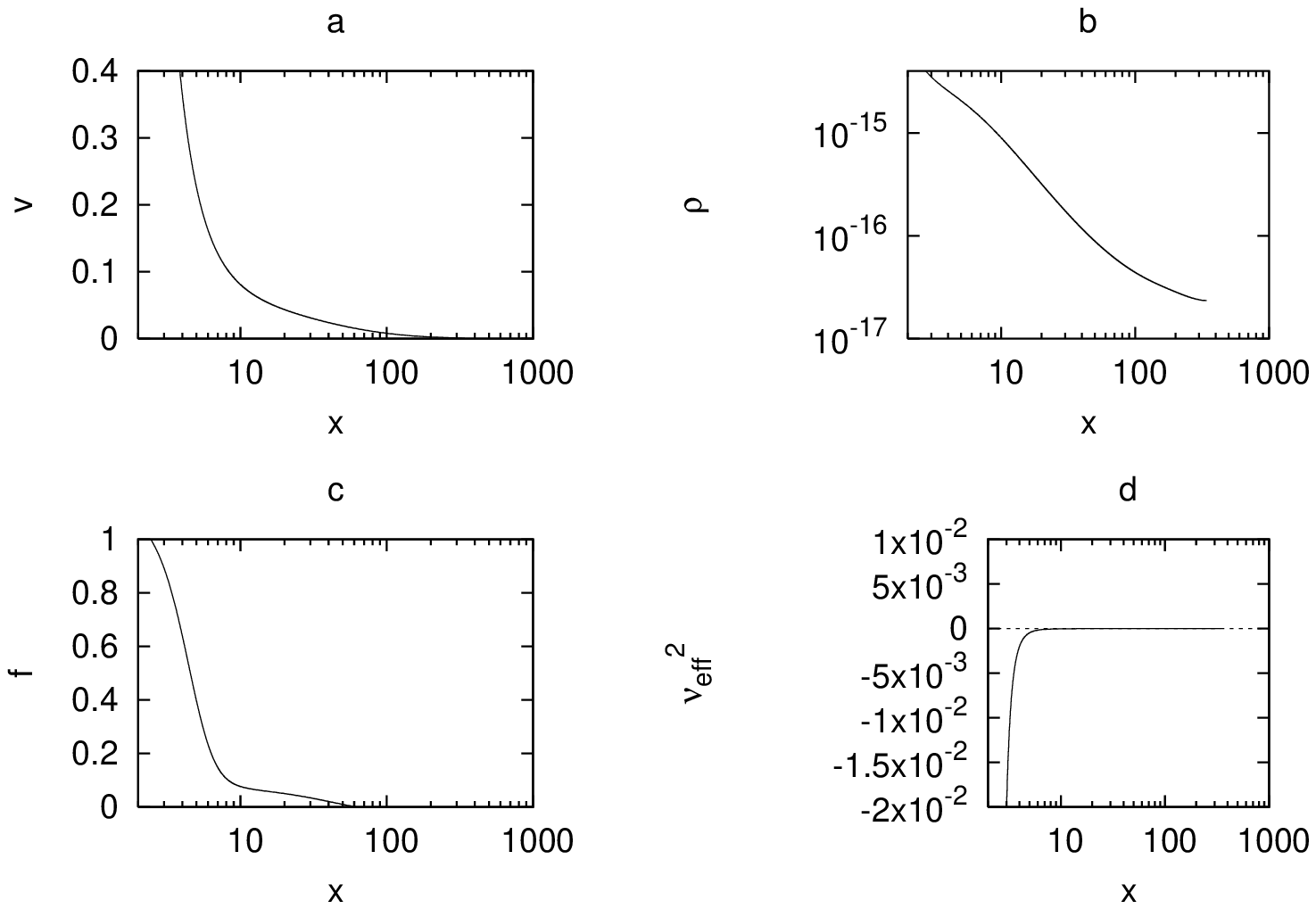}
\caption{
Variation of dimensionless (a) radial velocity, (b) density,
(c) cooling factor,
(d) square of convective frequency, as functions of radial
coordinate for $\mdot=10$.
Other parameters are $\alpha = 0.01$, $M=10^7$, $\gamma=1.345$; see Table 1 for details.
 }
\label{figsuh0}
\end{figure}

\begin{figure}
\centering
\includegraphics[angle=-90,width=0.90\columnwidth]{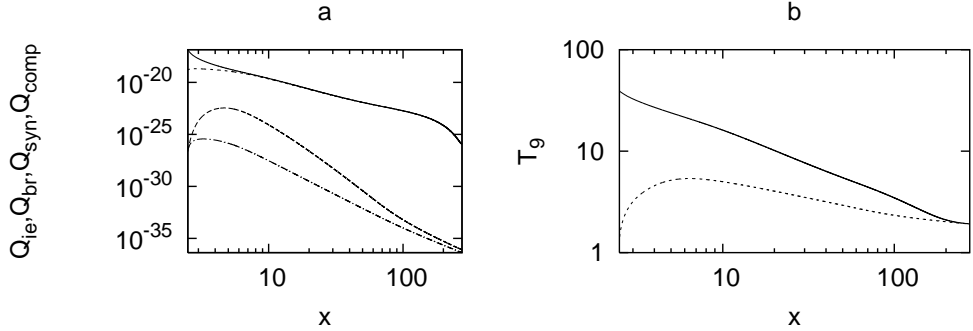}
\caption{
Variation of (a) dimensionless energy of Coulomb coupling (solid line),
bremsstrahlung (dotted line), synchrotron (dash-dotted line), inverse Comptonization due
to synchrotron photon (dashed line) processes in logarithmic scale,
(b) corresponding
ion (solid) and electron (dotted) temperatures in units of $10^9$K,
as functions of radial coordinate for $\mdot=10$.
Other parameters are $\alpha = 0.01$, $M=10^7$, $\gamma=1.345$; see Table 1 for details.
}
\label{figsuh0t}
\end{figure}

The basic flow properties (shown in Figs. \ref{figsuh0} and \ref{figsuh0t}) 
are pretty similar to those around stellar mass black holes, except 
that in the present case the centrifugal barrier smears out, as evident 
from Fig. \ref{figsuh0}a. This is due to high black hole mass causing 
the density of the flow to be low, resulting in a fast infall.
This also leads to, unlike that of a stellar mass black hole, an inefficient
synchrotron radiation even at the inner edge of the disc.
Similar to the stellar mass black hole, the flow transits from a
cooling dominated phase to a radiatively 
inefficient phase during infalling towards the black hole.

\section{Comparison between flows with different {\Large $\alpha$}
}

So far we have discussed models with a typical Shakura-Sunyaev
viscosity parameter $\alpha=0.01$.
Now we plan to explore a lower $\alpha$
to understand any significant change in the flow properties.

The rate of energy-momentum transfer between any two successive layers
of the fluid element naturally decreases for a lower value of $\alpha$,
which increases the residence time of the flow in the sub-Keplerian
disc. Moreover, due to inefficiency of the outward transport of 
angular momentum, a low $\alpha$ can not keep the flow Keplerian 
below a certain radius, resulting in a larger Keplerian-sub-Keplerian 
transition radius. Now, as we know, a high residence time of the 
flow in the disc corresponds to a high probability of cooling. 
Therefore, the flow with a lower $\alpha$ is expected to be
cooler with a small $f$. In order to compare, we consider the sub-Eddington
accretion with $\mdot=0.01$ around a supermassive black hole of mass
$M=10^7$, e.g. Sgr~$A^*$, with $\alpha=0.0001,0.01$.

\begin{figure}
\centering
\includegraphics[angle=-90,width=0.90\columnwidth]{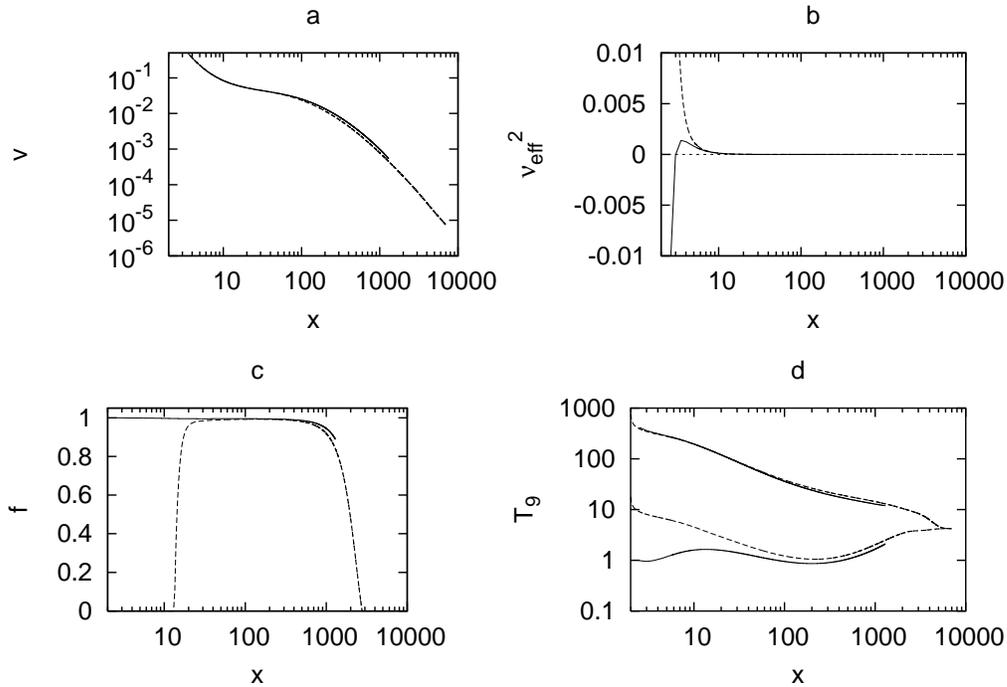}
\caption{
Comparison between solutions for high and low $\alpha$:
Variation of dimensionless (a) velocity, (b) square of convective frequency, (c) cooling factor,
(d) ion (upper set of lines) and electron (lower set of lines) temperatures,
a functions of radial coordinate, when solid lines correspond to
$\alpha=0.01$ and dashed lines correspond to $\alpha=0.0001$.
Other parameters are $\mdot=0.01$, $M=10^7$, $\gamma=1.5$, $a=0$.
}
\label{alfacom}
\end{figure}

Figure \ref{alfacom} shows that the velocity profile does not change
very significantly for $\alpha=0.0001$ compared to that of $\alpha=0.01$. 
However, importantly, the flow is in the cooling dominated single 
temperature phase at around the transition radius. As it advances fast in
the sub-Keplerian part, the efficiency of cooling decreases due to a 
decrease in the residence time of the flow in the disc rendering a 
transition to the radiatively inefficient phase, e.g. ADAF, with 
$f\rightarrow 1$.  Subsequently, at $x\le 17$, $f$ goes down again 
rapidly and reaches zero at $x\sim 10$ (see Fig. \ref{alfacom}c).
Hence, the transition from the radiatively inefficient ADAF phase to 
GAAF phase is very sharp. As a consequence the flow remains stable 
even upto the horizon (see Fig. \ref{alfacom}b). As the 
Keplerian-sub-Keplerian transition region for a low $\alpha$ is far 
away from the black hole, compared to that of a large $\alpha$, the 
gravitational effect is weaker there in the former case. Consequently, 
$T_e$ and $T_p$ merge before the flow reaches the transition region,
unlike in the case of a large $\alpha$ ($=0.01$) flow.

\section{Discussion and Summary}

We have investigated the two temperature accretion flow around black holes
with the self-consistent solutions of the complete set of hydrodynamic equations, 
appropriate for modeling disc flows, along with cooling processes. 
We have considered three important cooling processes:
bremsstrahlung, synchrotron and inverse-Comptonization due to synchrotron photons. 
Synchrotron emission is significant when the magnetic
field is high, which is particularly the case for a stellar mass black hole.

After solving the complete set of disc equations, we have seen, 
in several cases, that there is a transition in the flow 
from ADAF phase to GAAF phase and vice versa. This is
easily understood from the cooling efficiency factor $f$, calculated for
each model. Note  that we do not impose any restriction to the flow parameters,
unlike the previous authors (Narayan \& Yi 1994, 1995).
While the previous authors
especially restricted with flows having $f=1$, 
here we do not impose any such restriction
to start with and let the parameter $f$ be determined
self-consistently as the system evolves. Therefore, our model
is very general whose special case may be understood as an ADAF at a
particular region of the disc. 

\begin{figure}
\centering
\includegraphics[angle=-90,width=0.60\columnwidth]{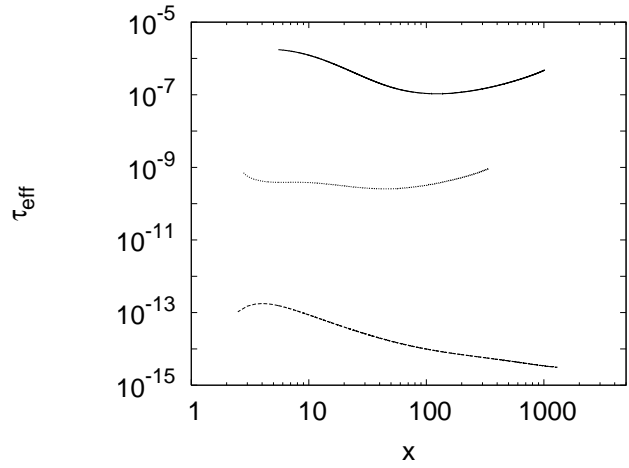}
\caption{
Variation of the effective optical depth as a function of radial coordinate
for the model cases given in Table 1. Solid and dotted curves
correspond to stellar and supermassive black holes with super-Eddington 
accretion rate and dashed curve corresponds to supermassive black hole with 
sub-Eddington accretion rate. See Table 1 for details.
}
\label{figtau}
\end{figure}

We have especially explored optically thin flows incorporating various 
nonthermal cooling processes. Figure \ref{figtau} shows the 
variation of the effective optical depth with disk radii for various
cases discussed here. Note that the maximum possible optical depth 
to be $\sim 10^{-6}$, supporting our choice of
optically thin flows strictly.

The present model can also explain the
under-luminous to ultra-luminous sources, stellar mass
to supermassive black holes. The luminosity of the under-luminous 
source Sgr~$A^{*}$ can be explained by a model with
$\mdot\lsim 0.0001$ and $M=10^6-10^7$. On the other hand
$\mdot\sim 100$ around a similar black hole can
explain the highly luminous AGNs like PKS~0743-67. The observed
luminosity of ULX sources can also be well fitted with this kind of high
accretion rate, for the stellar mass black hole. The computed 
luminosity is given in Table 2.

In general, a low mass accretion rate corresponds to a low density, which may
lead to weak emission processes with higher $f$. Hence, for a sub-Eddington
flow, $f$ is close to unity, giving radiatively inefficient ADAF, while
a super-Eddington flow may lead ultimately to a GAAF phase with $f<1$. However,
with a lower value of $\alpha$ the residence time of matter in the disc
increases, which further makes the disc act as a more efficient radiator.

In most of the cases, the ion and electron temperatures merge or
tend to merge at around the transition radius. This is because
the electrons and ions are in thermal
equilibrium and thus virial around the transition radius,
particularly at a high $\mdot$. As the sub-Keplerian flow advances,
the ions become hotter,
rendering the ion-electron Coulomb collisions weaker.
The electrons, however, cool down via processes like bremsstrahlung,
synchrotron emissions etc., keeping their temperature roughly constant
upto very inner part of the disc. This strictly reveals the two 
temperature nature throughout.

Next, one should try to predict spectra emitted from the accretion flow 
in the cases of different parameters based on the present solutions.
Naturally, unlike the optically thick Shakura-Sunyaev (1973) disk, the spectra
corresponding to GAAF should be dominated by non-thermal processes. This will 
provide important insights into the geometry and physics of
emitting regions. 

\section*{Acknowledgments}
This work is partly supported by a project, Grant No. SR/S2HEP12/2007, funded
by DST, India.
One of the authors (SRR) would like to thank the Council for Scientific and
Industrial Research (CSIR), Government of India, for providing a research 
fellowship.

\clearpage

\noindent{ Table 1: Parameters for $\alpha=0.01$, $a=0$,
when the subscript `c' indicates the quantity at the
sonic radius and $T_{ec}$ is expressed in units of $m_i c^2/k$ }

\begin{center}
\begin{tabular}{lllllllllllll}
\hline
\hline
$M$ & $\mdot$  & $\gamma$ & $x_{c}$ & $\lambda _{c}$ & $T_{ec}$  \\
\hline
\hline
$10$ & $10$   &  $1.345$& $5.5$ & $3.2$ & $0.000181565$\\
\hline
\hline
$10^7$ & $0.01$     & $1.5$  & $5.5$ & $3.2$ & $0.0001$ \\
\hline
\hline
$10^7$ & $10$       & $1.345$& $5.5$ & $3.2$ & $0.000427$\\
\hline
\hline
\end{tabular}
\end{center}


\centerline{ Table 2: Luminosity in erg/sec}

\begin{center}
\begin{tabular}{lllllllllllll}
\hline
\hline
 $M$ & $\mdot$ & $\gamma$ & $L$ \\
\hline
\hline
$10^7$       & $0.0001$ & $1.6$ & $10^{34}$ \\
$10^7$ & $100$  &  $1.34$ & $10^{47}$\\
$10$ & $100$   & $1.34$ & $10^{40}$\\
\hline
\hline
\end{tabular}
\end{center}

\end{document}